\documentclass{emulateapj}
\bibpunct{(}{)}{;}{a}{}{,}

\newcommand{\kap}[1]{Sect.\,\ref{#1}}
\newcommand{\jahre}{\ensuremath{\, \mathrm{yr}}}
\newcommand{\yr}{\ensuremath{\, \mathrm{yr}}}
\newcommand{\mem}[1]{\ensuremath{\mathrm{ #1}}}
\newcommand{\msun}{\ensuremath{\, {\rm M}_\odot}}
\newcommand{\lsun}{\ensuremath{\, {\rm L}_\odot}}

\newcommand{\czw}{\ensuremath{^{12}\mem{C}}}
\newcommand{\abb}[1]{Fig.\,\ref{#1}}
\newcommand{\teff}{\ensuremath{T_{\rm eff}}}

\newcommand{\n}{\newcommand}
\n{\bmain}{\begin{list}{${\bullet}$}{}}
\n{\emain}{\end{list}}
\n{\bminor}{\begin{list}{$\bf\triangleright$}{}}
\n{\eminor}{\end{list}}
\n{\Ms}{\mbox{$\,M_\odot$}}
\n{\Ls}{\mbox{$\,L_\odot$}}

\newcommand{\be}{\begin{eqnarray}}
\newcommand{\ee}{\end{eqnarray}}

\shorttitle{SN channel of SAGB Stars}
\shortauthors{Poelarends et al.\ (2007)}

\begin{document}

\title{The Supernova Channel of Super-AGB Stars}


\author{A.J.T.\ Poelarends\altaffilmark{1,2}, F.\
Herwig\altaffilmark{3,2}, N.\ Langer\altaffilmark{1}, and
A.\ Heger\altaffilmark{2,4,5}}

\altaffiltext{1}{Astronomical Institute, P.O. Box 80000, NL-3508 TA
Utrecht, The Netherlands, a.j.t.poelarends@astro.uu.nl, n.langer@astro.uu.nl}

\altaffiltext{2}{Theoretical Astrophysics Group, T-6, MS B227, Los Alamos National
Laboratory, Los Alamos, NM 87545, USA, aheger@lanl.gov}

\altaffiltext{3}{Keele Astrophysics Group, School of Physical and
  Geographical Sciences, Keele University, Staffordshire, UK, fherwig@astro.keele.ac.uk}

\altaffiltext{4}{Department of Astronomy and Astrophysics, University
of California, Santa Cruz, CA 95064, USA}

\altaffiltext{5}{Department of Astronomy and Astrophysics, University
of Chicago, 5640 S. Ellis Avenue, Chicago, IL 60637, USA}

\begin{abstract}
We study the late evolution of solar metallicity stars in the
transition region between white dwarf formation and core
collapse. This includes the super-asymptotic giant branch (super-AGB,
SAGB) stars, which have massive enough cores to ignite carbon burning
and form an oxygen-neon (ONe) core. The most massive SAGB stars have
cores that may grow to the Chandrasekhar mass because of continued
shell-burning.  Their cores collapse, triggering a so called electron
capture supernovae (ECSN).

From stellar evolution models we find that the initial mass range for
SAGB evolution is $7.5 ... 9.25 \msun$.  We perform calculations with
three different stellar evolution codes to investigate the sensitivity
of this mass range to some of the uncertainties in current stellar
models.  The mass range significantly depends on the treatment of
semiconvective mixing and convective overshooting. To consider the
effect of a large number of thermal pulses, as expected in SAGB stars,
we construct synthetic SAGB models that include a semi-analytical
treatment of dredge-up, hot-bottom burning, and thermal pulse
properties. To calibrate the time-dependent synthetic model, we have
calibrated a number of SAGB stellar evolution models. This synthetic
model enables us to compute the evolution of the main properties of
SAGB stars from the onset of thermal pulses until the core reaches the
Chandrasekhar mass or is uncovered by the stellar wind.
Thereby, we determine the stellar initial mass ranges that produce
ONe-white dwarfs and electron-capture supernovae.  The latter is found
to be $9.0 ... 9.25 \msun$ for our fiducial model, implying that
electron-capture supernovae would constitute about 4\% of all
supernovae in the local universe.  Our synthetic approach allows us to
explore the uncertainty of this number imposed by uncertainties in
the third dredge-up efficiency and ABG mass loss rate.  We find that
for both processes, the most optimistic approach leads to about a
doubling of the number of electron-capture supernovae, which provides
a firm upper limit to their contribution to all supernovae of
$\sim$20\%.
\end{abstract}


\section{Introduction}
It is well known that, for a given initial chemical composition, it is
the initial stellar mass which essentially determines the final fate
of a star: lower masses produce white dwarfs, higher masses neutron
stars and supernovae. The late evolution phases of stars in the
transition region between white dwarfs and neutron stars is
numerically difficult to model, and the relevant physics is not yet
fully understood. This mass range is therefore often omitted in
stellar evolution calculations. This is unsatisfactory because the
uncertain initial mass region for this evolution is $7$ to $12\msun$,
implying that as much as half of all supernovae may originate from
this transition region.

Of particular interest in this context is the evolution of so called
super-asymptotic giant branch (SAGB) stars, which ignite carbon
non-explosively, but also undergo thermal pulses
\citep{rgi96b,irg97,gri97b,rgi99,sie06}. These stars may end their
lives either as massive ONe-white dwarfs \citep{nom84e}, or as
electron capture supernovae (ECSNe), where the core collapse is
triggered by electron captures before Ne ignition \citep{wch98,
wti+03b}.  Stars of larger initial mass ignite hydrostatic neon
burning, form an iron core, and lead to classical core collapse
supernovae (CCSNe).

The upper mass limit of SAGB stars is affected by the second
dredge-up, which may occur after core He-exhaustion, and which
drastically reduces the mass of the helium core.  At this point and
throughout the following early SAGB phase carbon burning transforms
the CO core into an ONe core \citep{nom87c,rgi96b,irg97,gri97b,
rgi99}. Since the temperature is not high enough to ignite Ne, the
core cools, electron degeneracy in the core increases, and the
structure of the SAGB star outside of the ONe-core is very similar to
the most massive CO-core AGB stars \citep[for a general review of AGB
evolution see \citealt{ir83, hab96} and
\citealt{herwig:04c}]{frost:98b}. The degenerate core is surrounded by
the He- and H-shell sources, which eventually produce thermal pulses
due to the instability of the helium shell source \citep{yls04b}.

In this situation, the mass of the H-free core continues to grow. If
the core mass reaches the Chandrasekhar mass of $1.375\msun$, the core
will collapse triggered by electron captures on $^{24}$Mg and
$^{20}$Ne, and the star will become an ECSN
\citep{mny+80,mn87,hin93}. Recent studies by \citet{rgi96b, gri97b,
irg97, rgi99} and \citet{sie06} have shown that the mass fraction of
$^{24}\mathrm{Mg}$ in the ONe-core is smaller than previously thought,
which diminishes the role of electron captures on $^{24}\mathrm{Mg}$.
While \citet{gcg05} found that unburnt carbon in the degenerate ONe
core could trigger an explosion at densites of $\sim 10^{9}
\mathrm{g~cm}^{-3}$, we disregard this possibility furtheron since its
observational implications are not worked out, and therefore this
scenario can not yet be confronted with supernova observations.

The initial mass range for core collapse after SAGB evolution depends
on the effective core growth and mass loss of the SAGB star.  Larger
mass loss rates lead to a shorter duration of the SAGB phase. For very
high SAGB mass loss, there is no time for any significant core growth,
and the initial mass range for ECSNe will be very small. On the other
hand, the core growth rate in SAGB stars depends on the the hydrogen
shell burning and thus on two crucial factors, hot-bottom burning
\citep{sb91,vcs05}, and the efficiency of the third dredge-up.

Previous studies of SAGB stars have concentrated on the evolution of
the stellar cores \citep{nom84e, nom87c}. According to these models,
stars with helium cores between $2.0$ and $2.5\msun$\ form ONe cores and
explode as ECSN, leaving a neutron star less massive than $1.3\msun$.
\citet{rgi96b,rgi99}, \citet{irg97} and \citet{gri97b} studied the
evolution of complete SAGB stellar models in detail.  They describe
SAGB thermal pulses, and an outward mixing event which they called
dredge-out, in which the convective envelope connects to a convection
zone on top of the helium burning layer. \citet{sie06}, who studied
the effects of the carbon flame and of thermonuclear reactions on the
structure of the ONe core, finds similar results.

Thermal pulses in AGB evolutionary models require high numerical
resolution, both in time and space. The interpulse period decreases
with increasing core mass to eventually only a few years for the most
massive AGB star. In order to follow the evolution of SAGB stars with
very high core masses, orders of magnitude more thermal pulses (up to
10000s) have to be computed compared to low-mass AGB stars, which
experience only tens of thermal pulses. For this reason, no detailed
stellar evolution calculations through the entire super-TP-AGB phase
exist.  \citet{rgi99} attempted to characterize stars that would end
as ECSN.  Based on the assumption of a constant SAGB mass loss rate of
$10^{-4}\msun/\jahre$, they speculated that out of their set of four
calculated models ($9, 10, 10.5$ and $11\msun$) only the $11\msun$ model would
explode as an ECSN. The other models would lose all their envelope
before the core has grown enough, and their final fate would be an ONe
white dwarf. \citet{et04c} determine a minimum mass for supernova
explosion around $7\msun$ (with overshooting), or around $9\msun$
(without overshooting), again without being able to calculate the
stellar evolution models through the final phases.

Models of SAGB evolution suffer from two dominant sources of
uncertainty: mass loss and the efficiency of the third dredge-up.  To
explore these uncertainties would require to compute several model
grids, which is not feasible at this time.  We therefore take a
different approach and use the fact that TP-AGB stars, after a brief
transition phase, reach a quasi-steady state in which the important
structural quantities evolve in a simple and predictable way as
function of time.  This approach of synthetic AGB modeling has already
been successfully used for low-mass and massive AGB stars
\citep{rv81,ir83, marigo:96}.

In the following, we first describe the detailed stellar evolution
models (\kap{sect:method}) and identify the initial mass range for
SAGB stars by calculating the pre-AGB evolution phase up to the end of
the second dredge-up and possibly C-ignition, using three different
stellar evolution codes (\kap{sec:preAGB}). Next, we describe our SAGB
stellar evolution models (\kap{sect:sagb-evol}), and our synthetic
SAGB evolution model (\kap{sec:sagb-pop}). We present our results in
\kap{sec:results} and concluding remarks in \kap{sec:disc}.

\section{Numerical methods}
\label{sect:method}

We use three different stellar evolution codes to calculate the
evolution of solar metallicity stars up to the end of the second
dredge-up, or to Ne-ignition. We used the codes STERN
\citep{lan98,hlw00}, EVOL \citep{blo95,her00} and KEPLER \citep{wzw78,
hlw00}.  All three codes use the OPAL opacities \citep{ir96}, and are equipped
with up-to-date input physics, including a nuclear network with all
relevant thermonuclear reactions.

For our investigation, the most relevant difference between the codes
concerns the treatment of convective and semiconvective mixing.  As we
will see, these affect the He-core mass after central He-burning, and
thereby the final fate of the stellar model. STERN and KEPLER use the
Ledoux-criterion to determine convective instability, and take
semiconvection into account.  In KEPLER the treatment of
semiconvection leads to rather fast mixing.  Specifically, it is
approximated as a diffusive process with a diffusion coefficient that
is 10\,\% of the radiative diffusion coefficient.  Typically, this
leads to a 1000x shorter mixing time scale as for the default value of
\citet{langer:83} as used in STERN ($\alpha_{\mathrm{sem}}=0.01$).  No
modification to the temperature gradient is assumed, i.e., the
radiative temperature gradient is used.  Additionally, in KEPLER
convection zones are extended by one extra grid point where fast
mixing is assumed, to mimic convective overshooting.  In the EVOL
code, convective boundaries are determined by the Schwarzschild
criterion, and semiconvection is not treated as a separate mixing
process. Mixing beyond convective boundaries is treated by adopting an
exponentially decaying diffusion coefficient
\citep{herwig:97,her00}. Such mixing may be induced by convective
overshooting \citep{freytag:96}, or internal gravity waves
\citep{dt03}, or a combination of both \citep{young:05b}. For the
pre-AGB evolution, the overshoot parameter in EVOL has been set to
$f=0.016$, which was shown by \citet{her00} to reproduce the observed
main sequence width in the HR diagram of young open
clusters. Effectively, the strength of mixing in KEPLER lies in
between that of STERN (slow semiconvective mixing) and that of EVOL
(Schwarzschild criterion for convection is similar to very fast mixing
in semiconvective regions).

The EVOL code has previously been used to study low-mass
\citep[e.g.][]{herwig:04b} and massive AGB stars
\citep{her04b,herwig:04a}. KEPLER has in the past been applied to
study massive stars \citep{whw02}, but has not previously been used
for AGB simulations.  STERN has been used for low mass AGB stars
\citep{lhw+99,hll03,sgl04} as well as for massive stars \citep{hlw00,
hl00}.

\section{Pre-AGB evolution and the initial mass range for SAGB stars}
\label{sec:preAGB}
\begin{deluxetable}{lrrrrl}
\tablecaption{Summary of our detailed stellar evolution sequences. The
  columns give the model identifier (S means STERN, E is EVOL, and K
  is KEPLER), the initial mass ($\msun$), the helium core mass prior
  to the second dredge-up ($\msun$), the helium core mass after the
  second dredge-up ($\msun$), information the end of the simulation,
  and the final fate of the sequence according to our fiducial SAGB
  evolution properties (mass loss, dredge-up, as described in
  Sect.\,\ref{sect:pref_model}) }\label{tabl:overview} 
\tablehead{\colhead{Model} & \colhead{$M_{\rm i}$} &\colhead{pre-2DU}
  & \colhead{post-2DU} & \colhead{comments} & \colhead{fate}}
\startdata
S5.0 & 5.0 & 0.91 & 0.84 & 14 TP &CO WD\\
S8.5 & 8.5 & 1.73 & 1.02 & 10 TP &CO WD\\
S9   & 9   & 1.90 & 1.07 & 30 TP &ONe WD\\
S9.5 & 9.5 & 2.00 & 1.11 & &ONe WD\\
S10  & 10  & 2.14 & 1.16 & 55 TP & ONe WD\\
S10.5& 10.5& 2.30 & 1.20 & &ONe WD\\
S11  & 11  & 2.45 & 1.23 & &ONe WD\\
S11.5& 11.5& 2.61 & 1.27 & 15 TP &ONe WD\\
S12  & 12  & 2.79 & 1.32 & dredge-out &ECSN\\
S12.5& 12.5& 2.95 & 2.95 & dredge-out &CCSN\\
S13.0& 13  & 3.13 & 3.13 & Ne ignition & CCSN\\
S16.0& 16  & 4.33 & 4.33 & Ne ignition & CCSN\\
\hline
E6.5 & 6.5 & 1.59 & 0.99& &CO WD\\
E7.5 & 7.5 & 1.90 & 1.07& &ONe WD\\
E8.5 & 8.5 & 2.27 & 1.24& &ONe WD\\
E9.5 & 9.5 & 2.65 & 1.43& &CCSN\\
E10.0 & 10.0 & 2.82 & 2.82 & dredge-out &CCSN\\
E10.5 & 10.5 & 3.00 & 3.00 & Ne ignition &CCSN\\
\hline
E0099 & 9.0 & 2.15 & 1.17 & $f_{\mathrm{over}} = 0.004$ \\
\hline
K8   & 8.0 & 1.808 & 1.168& &ONe WD\cr
K8.5 & 8.5 & 1.955 & 1.247& &ONe WD\cr
K9   & 9.0 & 2.130 & 1.338& &ONe WD\cr
K9.1 & 9.1 & 2.161 & 1.357& &ECSN\cr
K9.2 & 9.2 & 2.190 & 1.548& Ne ignition &CCSN\cr
K9.3 & 9.3 & 2.221 & 1.603& Ne ignition &CCSN\cr
K9.4 & 9.4 & 2.253 & 1.690& Ne ignition &CCSN\cr
K9.5 & 9.5 & 2.283 & 1.799& Ne ignition &CCSN\cr
K10 & 10.0& 2.439 & 2.315& Ne ignition &CCSN\cr
K10.5 & 10.5& 2.598 & 2.596& Ne ignition &CCSN\cr
K11 & 11.0& 2.759 & 2.759& Ne ignition &CCSN\cr
\enddata
\end{deluxetable}
In order to identify the processes that lead to SAGB star formation we
calculate stellar evolution sequences with initial masses between
$6.5\msun$ and $13\msun$, starting from the zero age main sequence until
the completion of the second dredge-up or Ne ignition
(Table~\ref{tabl:overview}). Up to the end of the second dredge-up, no mass loss
is taken into account. The initial metallicity of our models is
$Z = 0.02$.  The effects of rotation or magnetic fields are not taken
into account.

\subsection{H- and He-core burning}
The evolution of stars toward the SAGB has been studied previously
\citep{rgi96b,irg97,gri97b,rgi99,sie06}, and our simulations
qualitatively confirm these results, although quantitative differences
occur.  In our STERN models, a consequence of including semiconvection
is that during core helium burning, a semiconvective layer limits the
mixing between the inner helium burning core and the outer convective
core, which still grows in mass (see also Fig.~\ref{fig:2dup} below).
This decreases the lifetime of the core helium burning phase, because
the available amount of helium is reduced, and leads to smaller helium
and CO-core masses compared to models which use the Schwarzschild
criterion for convection.

\citet{gbb+00} studied the effect of convective overshooting on the
maximum initial mass for which stars do not ignite carbon,
$M_{\mathrm{up}}$, and which defines the lower limit of SAGB
stars. They find for models without overshooting a value of
$M_{\mathrm{up}}$ of 6\msun ...7\msun, while a moderate amount of
overshooting reduces this by 1\msun. In our models we find
$M_{\mathrm{up}} = 7.5\msun$ (EVOL/KEPLER), while our STERN models --
without any overshooting -- give $M_{\mathrm{up}} = 9.0\msun$. We will
discuss these differences in the next paragraph.

\begin{figure}
\plotone{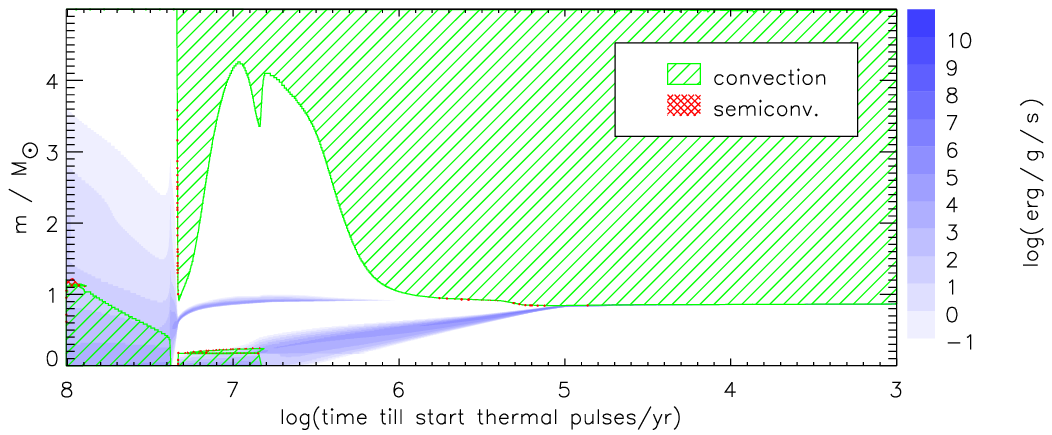}
\plotone{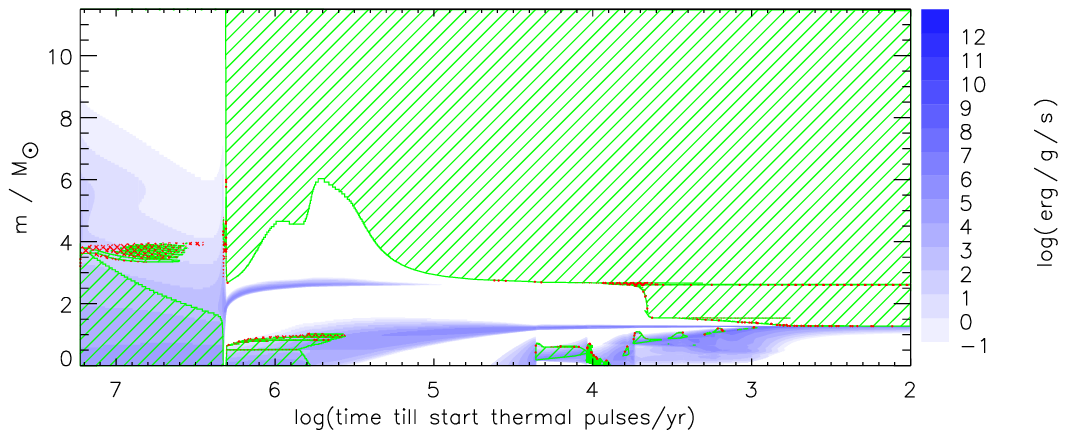}
\plotone{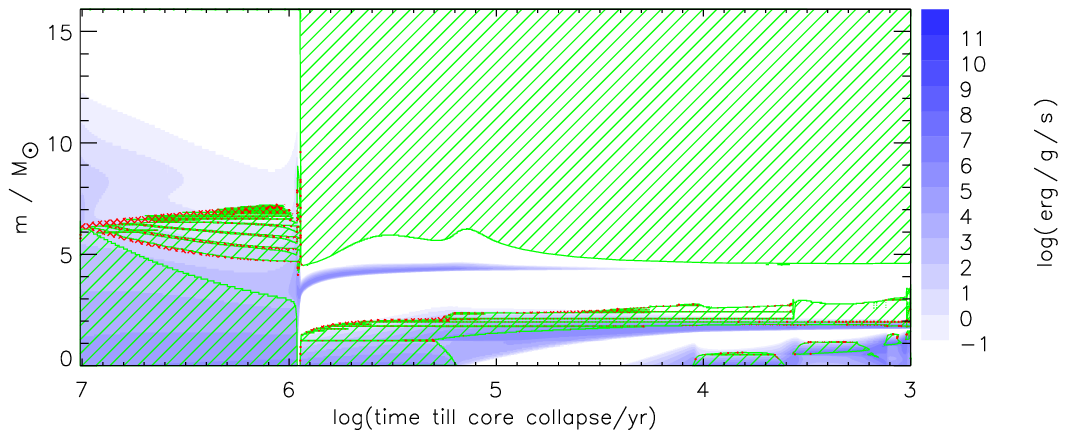}
\caption{Time evolution of convection zones and energy generation for
three evolution sequences with different mass, computed with
STERN. The initial masses and evolution scenarios are: top panel:
$5\msun$, massive AGB, middle panel:$11.5\msun$, SAGB, lower panel:
$16.0\msun$ star, Fe-core, CCSN. Convective regions are indicated by
green hatching while semiconvective regions are indicated by red
cross-hatching. The energy generation from nuclear burning is shown in
greyscale with a legend to the side in units of log erg
$\mathrm{g}^{-1}\ \mathrm{s}^{-1}$.}
\label{fig:2dup}
\end{figure}
\subsection{The second dredge-up}\label{sect:2DUP}
The second dredge-up is a key differences between SAGB stars and
massive stars that encounter Fe-core collapse. After core-He
exhaustion, the core resumes contraction while the envelope
expands. As the star evolves up the asymptotic giant branch the
envelope convection deepens, and eventually penetrates into the H-free
core.  Only due to this mixing event is the H-free core mass
sufficiently reduced so that an electron-degenerate core can form
which then cools and prevents Ne from igniting.  If the core mass after
the 2nd dredge-up is smaller than the Chandrasekhar mass, an
electron-degenerate core will form and the He- and H-shells will
eventually start the thermal pulse cycle.

The dependence of the late evolutionary phases, including the second
dredge-up, on the initial mass is illustrated in the
Kippenhahn-diagrams of three sequences computed with the STERN code
shown in Figure~\ref{fig:2dup}. All models evolved through core-H and
core-He burning. In the $5.0\msun$ models, the hydrogen burning
terminates, and the second dredge-up reduces the helium core mass by
about $0.2\msun$. When the helium shell source gets close to the
bottom of the convective envelope, hydrogen reignites, and the thermal
pulse cycle starts.  For the $11.5\msun$ model, central hydrogen and
helium burning is followed by off-center carbon ignition. During the
carbon burning phase the size of the helium core is reduced by a deep
second dredge-up, after which the core becomes degenerate and thermal
pulses develop. In the $16.0\msun$ case, convective core H- and
He-burning is followed by core C-burning, and no 2nd dredge-up occurs.
Ne ignites hydrostatically, and subsequent burning will lead to the
formation of an iron core.

In accord with previous work \citep{rgi96b,irg97,gri97b,rgi99,sie06},
the second dredge-up reduces the helium core mass to values below the
Chandrasekhar mass in our EVOL and STERN models. This leads to a clear
definition of the upper mass limit of SAGB stars, as the critical mass
between the occurrence and non-occurrence of the second dredge-up. For
initial masses lower than this critical mass, Ne ignition is always
avoided, while for initial masses higher than the critical mass the
helium core mass is so large ($\sim 2.8\msun$) that Ne always
ignites. Our KEPLER models show a more complicated behavior: some show
a second dredge-up depth which leaves helium cores with masses in
between $1.4\msun$ and $2.8\msun$. However, those with post-dredge-up
helium cores above the Chandrasekhar limit all ignite core Ne
burning. We conclude that a second dredge-up down to the Chandrasekhar
mass is required for an ECSN to occur, which thus defines our upper
SAGB mass limit. This is also in line with the recent results of
\citet{et04}, who do indeed find Ne shell flashes in some of their
most massive models undergoing the second dredge-up; however the
dredge-up does proceed to the Chandrasekhar mass, and the suggested
fate of these models is that of an ECSN.

\begin{figure}
\includegraphics[angle=270,scale=0.31]{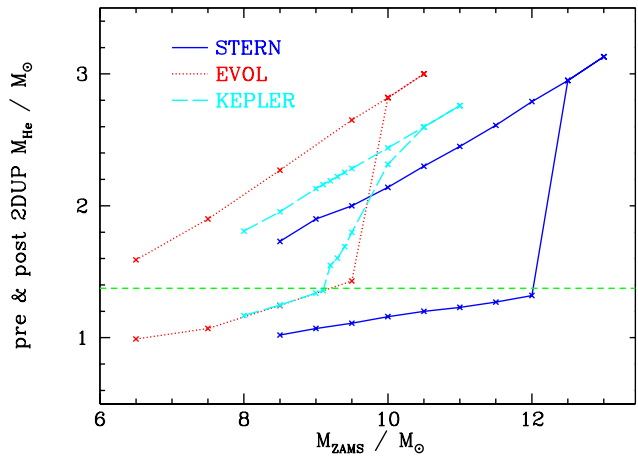}
\caption{Helium core masses for stars of various initial masses, as
obtained using different stellar evolution codes (solid line: STERN,
dashed line: KEPLER, dotted line: EVOL).  The upper part of the line
shows the maximum size of the helium core, prior to the second
dredge-up. The lower part shows the size of the helium core just after
the completion of the second dredge-up and prior to the onset of the
TP-AGB. The light dashed horizontal line gives the lower limit for the
final helium core mass for which the star may experience an
electron-capture supernova. }
\label{fig:core}
\end{figure}

Figure~\ref{fig:core} shows the helium core masses obtained in our
detailed stellar evolution models. While the core mass after the 2nd
dredge-up increases with initial mass for models computed with all
three codes, differences arise with respect to the critical mass for
second dredge-up. KEPLER and EVOL have similar final core masses,
however they differ slightly with respect to the maximum core
masses. These differences are related to the treatment of convection
and overshooting. STERN uses the Ledoux-criterion for determining the
convective boundaries, which naturally gives rise to smaller cores
than the Schwarzschild-criterion. In these models no rotation was
included, which -- if included -- would give significantly larger
cores, due to rotationally induced mixing during the hydrogen and
helium burning phases \citep{hlw00}.

Using EVOL and KEPLER we find the transition between stars with and
without a deep second dredge up at $\sim 9.25\msun$. On the other
hand, using STERN we find that stars more massive than $12\msun$ do
not experience a deep second dredge-up. The models in the mass range
between $12\msun$ and $12.5\msun$ show a convective shell that
develops on top of the helium burning layer (the so called dredge-out,
c.f. \citealt{rgi99}) which connects through a semiconvective layer
with the bottom of the hydrogen-rich convective envelope. We find that
protons are mixed into this hot layer and burn quickly. For a proper
treatment of this interaction a scheme that solves the burning and
mixing simultaneously is needed to follow the subsequent evolution of
these stars. Since STERN is not equipped with such a scheme but
calculates burning and mixing separately, we followed the subsequent
evolution with very small timesteps until the code was not able to
calculate further. Therefore, it remains unclear whether this
semiconvective layer dissolves and on what time scale. If it would,
the helium core masses would be reduced to just below the
Chandrasekhar mass. If the semiconvective layer remains for the rest
of the evolution of the star, it would allow Ne to ignite in the core
and eventually lead to Fe-core collapse supernova. This renders the
upper mass limit for SAGB stars according to the STERN models somewhat
ambiguous in the range $12...12.5\msun$.

In the EVOL models, the convective core overshooting was calibrated by
the observed width of the main sequence. However, stars rotate, and
the STERN code usually takes this into account. The effect of rotation
also widens the main-sequence and in this way STERN models with
rotation can reproduce the observed main-sequence width as well
\citep{hl00}.  In this study we compute non-rotating models with
STERN, in order to avoid the complex question of how rotational mixing
affects SAGB properties. As a drawback, the initial mass range for
SAGB stars found with STERN is offset compared to the EVOL/KEPLER
models by $\approx 2.75\msun$.  Since the non-rotating STERN models do
not reproduce the well-established main-sequence width, we prefer here
the results of the EVOL/KEPLER models to derive the initial mass range
for SAGB stars.  Based on those, the upper mass limit for SAGB stars
-- and thus electron capture supernovae -- is about $9.25\msun$
(Figure~\ref{fig:core}).  Without knowledge about the subsequent phase
-- the thermally pulsing SAGB phase -- we can only say that the lower
limit for electron capture supernovae is $7.5\msun$, since stars with
a lower mass do not ignite carbon.

\section{The TP-SAGB stellar evolution models}\label{sect:sagb-evol}

\begin{figure}
\plotone{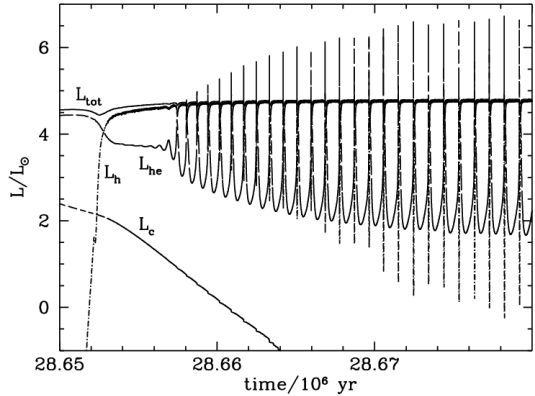}
\caption{Nuclear burning luminosity contributions and total luminosity
   as a function of time at the onset of the thermal pulse phase for a
   $9\msun$ star calculated with STERN. At the time of second
   dredge-up hot-bottom burning starts, shown in the figure as a steep
   rise of the hydrogen luminosity. }
\label{fig:lumplot}
\end{figure}

\subsection{Thermal Pulses and Hot Bottom Burning}\label{sect:TP}
Double shell burning of H and He on degenerate cores leads to periodic
thermonuclear instabilities. These He-shell flashes or thermal pulses
are an important site for nucleosynthesis in AGB stars, and cause
mixing of the intershell region and --- by way of the third dredge-up
--- mixing of processed material to the surface
\citep{ir83,busso:99,herwig:04c}. Thermal pulses of SAGB stars are
similar to thermal pulses of CO-core AGB stars \citep{rgi96b}. In
order to obtain quantitative information on these SAGB thermal pulse cycles,
we calculate such model sequences for several initial masses
(Table~1).

As in massive AGB stars, most of the luminosity is produced by
hot-bottom burning. During hot-bottom burning, hydrogen is transported
convectively into the H-shell, and H-burning ashes are transported out
of the shell into the envelope.  In the more massive SAGB stars this
hot-bottom burning starts immediately after the completion of the
second dredge-up, and can proceed at very high temperatures. In our
STERN models we obtain values of $1.0 \times 10^{8}\mathrm{K}$
($10\msun$ with $M_{\rm c}=1.16\msun$ after 30 thermal pulses) and
$1.1 \times 10^{8}\mathrm{K}$ ($11.5\msun$ with $M_{\rm c}=1.27\msun$
already at the first thermal pulse). The EVOL models show a similar
trend with the $9.0\msun$ model (E0099) reaching temperatures at the
bottom of the convective envelope of $1.13 \times 10^{8}\mathrm{K}$
after the 12th pulse.

Hot-bottom burning could be stronger (or weaker) than in our
calculations, e.g.\ due to convective overshooting at the bottom of
the convective envelope, or due to a larger (or smaller) convective
efficiency than assumed in most MLT based stellar evolution
calculations. In that case, the accretion of He on the core may be so
much reduced that the core does not or only very slowly grow. We have
performed some test calculations with enhanced convective extra mixing
during the hot-bottom phase. These tests show so far a stationary H-shell,
which slowly processes its envelope, and no core growth, which results
probably in a massive ONe white dwarf. Whether this theoretical
possibility is occuring in real stars is not clear because the physics
of a convective boundary inside the H-shell is poorly known.

SAGB stars show a He-peak luminosity during thermal pulses of around
$\log L/\lsun \sim 6$, which is significantly lower than obtained in
massive AGB stars which reach luminosities up to $\log L/\lsun \sim
8$. This may explain why the third dredge-up is less efficient in
terms of the dredge-up parameter $\lambda$ (see
Sect.\,\ref{sect:3dup}).

Extending the trend seen from low-mass to massive AGB stars, SAGB
stars have smaller intershell masses (in the STERN $9\msun$ model of
$7\times 10^{-4}\msun$ at a core mass of $1.06\msun$), and the
interpulse time is also lower, ranging from $50 \yr$ for the
$11.5\msun$ SAGB star with core mass of $1.27\msun$ to $1000 \yr$ for
a SAGB star with a mass of $9.0\msun$.

\subsection{Efficiency of the 3rd dredge-up}
\label{sect:3dup}
The growth of the core during the TP-SAGB may be decreased by the
dredge-up of material after a thermal pulse. The efficiency of the
dredge-up (DUP) is expressed through the dredge-up parameter $\lambda=\Delta
M_{\mathrm{H}} / \Delta M_{\mathrm{DUP}}$, where $\Delta
M_{\mathrm{H}}$ is the core mass increase due to H-burning during the
interpulse phase, and $\Delta M_{\mathrm{DUP}}$ is the mass that is
dredged up by the convective envelope.

In the models calculated with STERN we did not observe any
dredge-up. This result is consistent with results for non-rotating low
mass AGB-stars \citep{sgl04}
from the same code which are also calculated using the
Ledoux-criterion for convection.  \citet{rgi96b} and \citet{sp06} find
a similar result.  The recent models of \citet{dl06} find very
efficient dredge-up, e.g.\ $\lambda \approx 0.7$ for a $9.5\msun$
model.  Observations clearly require a 3rd dredge-up in low-mass and
massive AGB stars, since we see its result in terms of carbon and
s-process enrichment in real AGB stars.  However, the efficiency of
the 3rd dredge-up in massive and SAGB stars is not constrained
observationally, partly probably because of the high dilution factor
in the massive envelope.

In order to get an idea about the efficiency of the 3rd dredge-up in
super AGB stars and the robustness of our and previous results, we
studied the behavior of the thermal pulses also with the EVOL code. We
calculated a $9\msun$ model (E0099) until the 12th pulse. This model
was computed with a four times smaller factor for the overshooting
than the other EVOL models ($f_{\mathrm{over}} = 0.004$) until the
TP-AGB. This gives the star a smaller core than the regular models. On
the TP-AGB a value of $f_{\mathrm{over}} = 0.008$ was used. The first
thermal pulse starts after the completion of the second dredge-up,
when the bottom of the convective envelope is at $m_{\mathrm{r}} =
1.17\msun$. The surface luminosity after 12 pulses is $\log
\mathrm{L/\Ls} = 5.07$, the maximum helium luminosity during the 12th
pulse $\log \mathrm{L_{\mathrm{He}}/\Ls} = 6.17$, and the duration of
the interpulse period is $\sim 500 \mathrm{yr}$.

\begin{figure}
\plotone{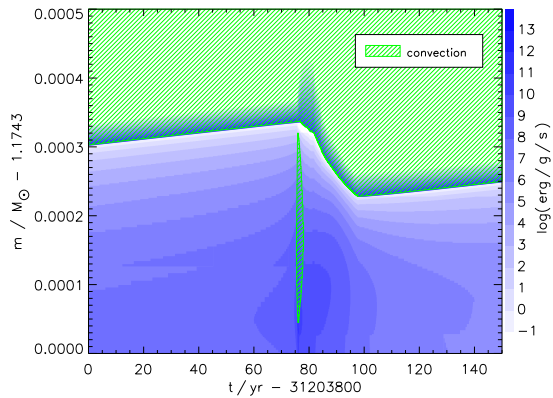}
\caption{Evolution of the SAGB He-shell flash during the 12th pulse
computed with the EVOL code (E0099). The dredge-up efficiency equals
$\lambda = 0.5$. }
\label{fig:hotdup}
\end{figure}

After the eighth pulse, the ensuing mixing has the characteristics of
a 'hot' dredge-up, first described for massive low-metallicity AGB
stars by \citet{her04b} and also found by \citet{cdl+01} for $Z=0$
models. Any small amount of mixing of protons into the hot \czw-rich
layers --- performed here by diffusive exponential overshooting ---
leads to violent H-burning which increases the convective
instability. Like a flame, this corrosive hydrogen burning enforces
the penetration of the convective envelope into the intershell (see
Fig.~\ref{fig:hotdup}).  For this situation, we find efficient
dredge-up ($\lambda \sim 0.5$), i.e. half of the interpulse core
growth is dredged up, reducing the average pulse cycle core growth
rate. Unfortunately the hot dredge-up phenomenon adds another source
of uncertainty to dredge-up predictions as the dredge-up efficiency is
extremely sensitive to the overshooting or extra-mixing efficiency at
the bottom of the convective boundary.
 
\section{The SAGB population synthesis model}\label{sec:sagb-pop}
Mass loss and the dredge-up are the two most important but also most
uncertain processes that determine the final evolution of SAGB
stars.  
Here we employ a simplified synthetic model that allows us to estimate
the effect of different assumptions concerning these two processes
on the initial mass range for ECSNe.

\subsection{A simple estimate}
\label{sec:sagb-simp_pop}

\begin{figure}
\includegraphics[angle=270,scale=0.31]{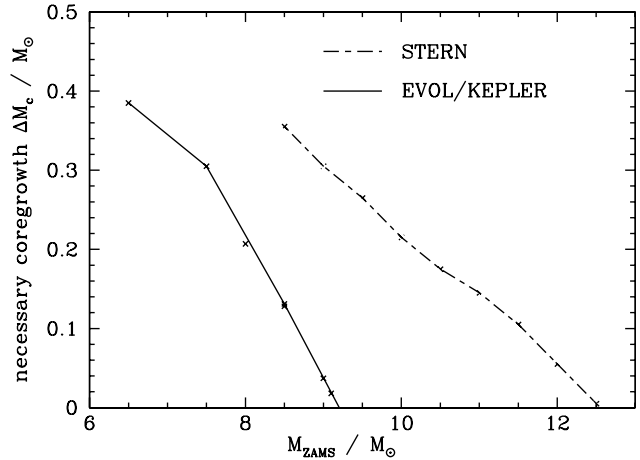}
\caption{Mass $\Delta M_{\mathrm{c}}$ by which the core needs to grow
during the TP-AGB in order to reach the Chandrasekhar mass, as
function of the initial mass.
The upper boundary is given by stars that do not have a second
dredge-up, so their cores are larger than the Chandrasekhar mass.  }
\label{fig:necce_massloss}
\end{figure}

We start with a simple back-of-the-envelope estimate: Stars that have,
after carbon burning, 
a helium core mass larger than the Chandrasekhar mass
($M_{\mathrm{Ch}}$) explode as CCSN.  The Chandrasekhar
mass of a cold iron core is $M_{\mathrm{Ch-eff}}= 1.375\msun$
 \citep{sn80, nom81}.  In order to form an ECSN, the core mass has to
grow from the mass at the beginning of the TP-AGB, $M_{\mathrm{c}}(2DUP)$,
to the Chandrasekhar mass by 
\be 
\Delta
M_{\mathrm{c}}=M_{\mathrm{Ch}}-M_{\mathrm{c}}(2DUP) .  
\ee 
This value depends strongly on the initial mass as
Figure~\ref{fig:necce_massloss} shows.

Whether the core is able to grow by this amount depends only on the
mass of the envelope, the core growth rate and the mass loss rate.
Given these quantities, $\Delta M_{\mathrm{c,max}}$ is the maximum
mass that the core can grow. 

\begin{figure}
\includegraphics[angle=270,scale=0.31]{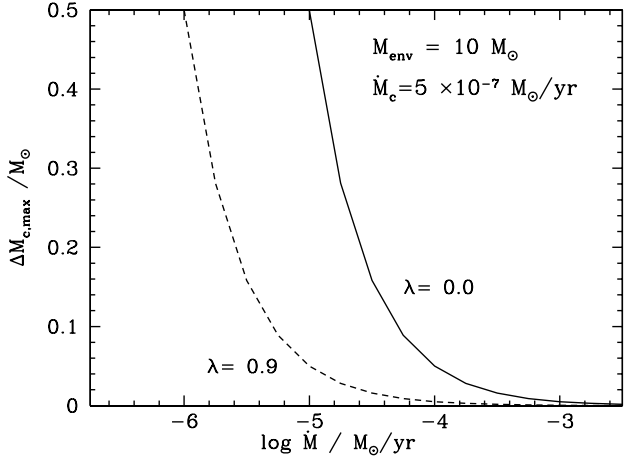}
\caption{Expected core growth during the TP-SAGB
(Eq.~\ref{eq:core_growth}) as function of the mass loss rate, for two
different dredge-up efficiencies $\lambda$ as labeled.  A constant
mass loss rate, an envelope mass of $10\msun$, and a constant core
growth rate of $5\times 10^{-7}\msun/\mathrm{yr}^{-1}$ are assumed. }
\label{fig:massloss}
\end{figure}

The core growth rate due to nuclear burning is
$\mathrm{d}M_{\mathrm{c}}/\mathrm{d}t$.  Due to the 3rd dredge-up, the
value for the core growth rate can decrease. We correct for this by
introducing a factor $1-\lambda$. The timescale on which the envelope of
the star will be lost is 
\be
  \tau_{\mathrm{env}}=\frac{M_{\mathrm{env}}}{\mathrm{d}M/\mathrm{d}t_{\mathrm{env}}}, 
\ee
and multiplied by the core growth rate this gives the maximum value
that the core can grow. This gives an approximate relation
for the growth of the core during the TP-SAGB phase
\be\label{eq:core_growth} \Delta
M_{\mathrm{c,max}}=\frac{M_{\mathrm{env}}}{\mathrm{d}M/\mathrm{d}t_{\mathrm{env}}}
\times (1-\lambda)\mathrm{d}M_{\mathrm{c}}/\mathrm{d}t .  \ee
For a typical, but constant, core growth rate of
$\dot{M}_{\mathrm{c}}= 5 \cdot 10^{-7}\msun \mathrm{yr}^{-1}$, and an
envelope mass of $M_{\mathrm{env}}=10\msun$, Fig.~\ref{fig:massloss}
shows $\Delta M_{\mathrm{c,max}}$ as a function of the mass loss rate,
for two different values of $\lambda$ (no dredge-up and $\lambda =
0.9$).

Figure~\ref{fig:necce_massloss} shows that in order to have an initial
mass range for ECSN of, for example, 1\msun, the core growth during
the SAGB phase must be of the order 0.1-0.2\msun.
Figure~\ref{fig:massloss} shows that if SAGB mass loss is larger than
$\approx 10^{-4}\msun/\mathrm{yr}$ such a core growth can not be
achieved, even for inefficient 3rd dredge-up. For mass loss rates
below $\approx 10^{-6}\msun/\mathrm{yr}$, however, a core growth of a
few $0.1\,\msun$ is predicted even if $\lambda=0.9$.  Compared with
the empirical mass loss rates derived by \citet[hereafter L05,
c.f.\ Table~\ref{tabl:massloss}]{lcz+05} it is clear that the intial
mass range for ECSN is sensitive to the third dredge-up.

\subsection{Synthetic SAGB evolution}\label{sec:sagb-synt}
A quantitative estimate of the initial mass range for ECSN can be
obtained through a synthetic model for the TP-AGB phase, similar to
that of \citet[hereafter I04]{itk+04} for AGB stars, which is based on
detailed AGB models from Karakas \citep{klp02}. The extension to SAGB
stars is made by fitting the TP-AGB evolution of detailed stellar
evolution models (STERN) presented above, specifically over the mass
range between $7$ and $11.5\msun$ in initial mass.

Based on the SAGB STERN evolution sequences with up to 30 thermal
pulses, we derive fits for luminosity, radius, and Q-factor (see
Sect.~\ref{sect:hbb}), as function of the core mass
($M_{\mathrm{c}}$), the envelope mass ($M_{\mathrm{env}}$) and as
secondary parameters the metallicity ($Z$) and the envelope hydrogen
abundance ($X_{\mathrm{H}}$). Since the SAGB evolution models have
entered into a quasi-steady state regime, these fits are good
approximations for the subsequent evolution of SAGB stars during the
TP-AGB in mass (total, core and envelope), luminosity and radius.

We then use these analytic expression as basis for our synthetic
TP-SAGB model. As starting values for our synthetic SAGB calculation
we use total mass, core mass, and envelope hydrogen abundance after
the second dredge-up. First the luminosity is calculated from the
initial core and envelope mass, then the radius is calculated, which
is a function of the previously calculated luminosity and the envelope
mass, and finally the core growth is calculated and integrated over a
timestep $\mathrm{d}t$. From these quantities, the effective
temperature, mass loss rate, the resulting new mass of the envelope,
and the new mass of the core are calculated. The new core mass and
envelope mass are used as input for the next timestep.

In following subsections we describe the basic outline of our
synthetic model (for details we refer to \citealt{itk+04}).

\subsubsection{Luminosity and Radius}
We follow I04 and model the luminosity with two terms, one which
contains a core-mass--luminosity relation (CMLR) and one term due to
hot-bottom burning. The total luminosity of the star can now be
written as (cf. their Eq.~29)
\be
L=f_{\mathrm{d}}(f_{\mathrm{t}}L_{\mathrm{CMLR}}+L_{\mathrm{env}})\lsun,
\ee 
where $L_{\mathrm{CMLR}}$ is the core mass-luminosity relation given
by
\be
  L_{\mathrm{CMLR}}&=3.7311\, 10^4 \hfill \times\nonumber\\ 
  &  \mathrm{max}\lbrack (M_{\mathrm{c}}/\msun - 0.52629)(2.7812-
  M_{\mathrm{c}}/\msun),\nonumber\\ 
  & 1.2(M_{\mathrm{c}}/\msun -0.48)
  \rbrack  
\ee
if the core mass at the first thermal pulse, $M_{\mathrm{c,1TP}}$ 
is  $\geq 0.58\msun$.
%
%

$L_{\mathrm{env}}$ is the contribution due to hot-bottom burning
(e.g., I04:32),
\be
&L_{\mathrm{env}}&=1.50\, 10^4 \left(\frac{M_{\mathrm{env}}}{\msun}\right)^{1.3}\left[1 
+0.75\left(1-\frac{Z}{0.02}\right)\right]\nonumber\\
&\times& \mathrm{max}\left[\left(\frac{M_{\mathrm{c}}}{\msun} + 
\frac{1}{2}\frac{\Delta M_{\mathrm{c,nodup}}}{\msun} -0.75\right)^2,0\right].
\ee
$M_{\mathrm{c}}$ is the core mass, $M_{\mathrm{env}}$ is the envelope
mass. $\Delta M_{\mathrm{c,nodup}}$ is the change in core mass without
third dredge-up and is defined by
%
$\Delta M_{\mathrm{c,nodup}}=M_{\mathrm{c,nodup}}-M_{\mathrm{c,1TP}} $
%
with $M_{\mathrm{c,nodup}}$ the core mass as if there was no third
dredge-up and $M_{\mathrm{c,1TP}}$ the core mass at the first thermal
pulse. $Z$ is the metallicity. Note that we use a lower exponent
than I04 in the contribution of the envelope mass, i.e. 1.3 instead of
2, which resulted in good fits for models between $7$ and $11.5\msun$.

The function
\be
f_{\mathrm{t}} = \mathrm{min} \left[\left(\frac{\Delta M_{\mathrm{c,nodup}}/\msun}{0.04}\right)^{0.2},1.0\right]
\ee
accounts for the steep rise in luminosity at the beginning of the
TP-AGB. The function
\be
f_{\mathrm{d}}=1-0.2180\mathrm{exp}\left[-11.613(M_\mathrm{c}/\msun-0.56189)\right]
\ee
corrects for the short timescale dips in the luminosity during the thermal pulse
cycle.

For the fit to the radius we use an expression of the same form as
given by I04, but with coefficients adjusted to the STERN models:
\be 
\log(f_{\mathrm{r}} R) = -0.26 + 0.75\log\left(L/\lsun\right) - 0.41\log
M_{\mathrm{env}}/\msun
\ee
with
\be
f_{\mathrm{r}}= 0.09\log\left(M_{\mathrm{env}}/M_{\mathrm
  {env,0}}\right)
\ee 
a factor that accounts for the removal of the envelope, where
$M_{\mathrm{env,0}}$ is the mass of the envelope at the first thermal
pulse.  This correction factor is determined by a fit to a $9\msun$
model to which an extreme mass loss rate of $10^{-3}
\msun/\mathrm{yr}$ was applied.  For envelope masses below
$M_{\mathrm{env}} = 2\msun$ the fit predicts too large radii and it is
not valid for temperatures below $2500\,$K as calculated from $R$ and
$L$.

\subsubsection{Third Dredge-up}
\label{sect:lambda}

\begin{figure}
\includegraphics[angle=270,scale=0.31]{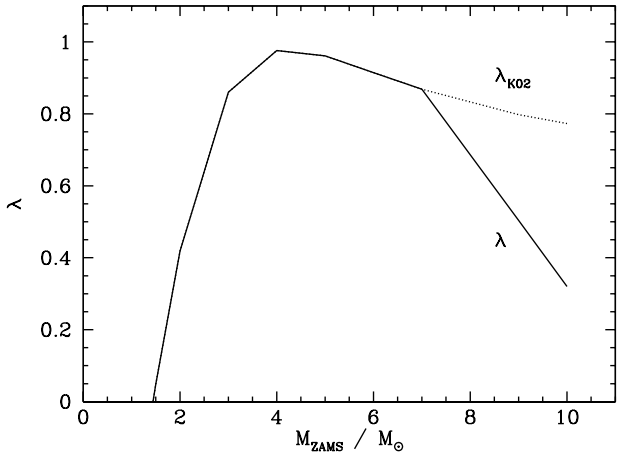}
\caption{Dredge-up efficiency as a function of initial mass. The
dotted line extrapolates the data by \citet{klp02}, while the solid
line gives the modification based on our SAGB dredge-up stellar
evolution sequence.}
\label{fig:3DUP}
\end{figure}

For the dependence of the third dredge-up on the initial mass we use
the data from \citet{klp02}. Our own EVOL SAGB models, however, show
smaller dredge-up ($\lambda = 0.5$ for $M_\mem{ini}=9 \msun$) than the
extrapolation of the \citet{klp02} data (Figure \ref{fig:3DUP}). We
therefore extend the fit to higher masses with a relation that
reflects our own data at $M_\mem{ini}=9 \msun$. To simulate a
situation with no dredge-up, we also include in our synthetic code an
option to set $\lambda = 0$. 

\subsubsection{Core growth and hot-bottom burning}\label{sect:hbb}

The growth rate of the He core in the inter-pulse phase is given by
\be 
\frac{\mathrm{d}M_{c}}{\mathrm{d}t}=Q \times L 
\ee 
where $L$ is the total luminosity of the star, and $Q$ gives the mass
of nuclear ashes accreted onto the core per energy released by the
star.  $Q$ depends on several model properties, especially on the
hot-bottom burning efficiency, and less on the chemical composition of
the envelope.  For massive AGB and SAGB stars its strength depends on
the envelope mass.  If the hot-bottom burning is efficient, $Q$ can be
small (see discussion in \S~\ref{sect:TP}), and the core may not
significantly grow at all.  \abb{fig:Q_fact} shows the decrease of $Q$
with increasing envelope mass.  We parameterize $Q$ as
\be 
Q=\mathrm{min}\lbrack 1.43\times 10^{-11}, 1.40 \times
10^{-11}+ \nonumber\\
\frac{4.166\times 10^{-12}}{X_{\mathrm{H}}}-
1.5\times10^{-12}\frac{M_{\mathrm{env}}}{\msun}\rbrack.  
\ee 
This parameterization is in reasonable agreement with I04 who
set Q to $1.585 \times 10^{-11} \msun\lsun^{-1}\mathrm{yr}^{-1}$,
\citet{hpt00} who found $1.27\times
10^{-11}\msun\lsun^{-1}\mathrm{yr}^{-1}$, and $1.02\times
10^{-11}\msun\lsun^{-1}\mathrm{yr}^{-1}$ in \citet{wg98}.

\begin{figure}
\includegraphics[angle=270,scale=0.31]{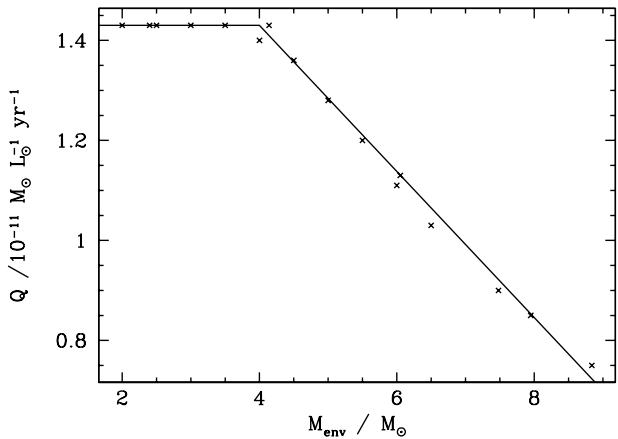}
\caption{Ashes accreted on core per unit stellar energy release ($Q$)
as a function of the mass of the envelope.  For envelopes less massive
than of $4\,\msun$ there is no hot-bottom burning.}
\label{fig:Q_fact}
\end{figure}

\subsubsection{Mass loss}
\label{sect:massloss}

As discussed in \S~\ref{sec:sagb-simp_pop} the mass loss of SAGB stars
is the other important factor to determine the initial mass range of
ECSN.  SAGB stars are O-rich (because of hot-bottom burning), and have
stellar parameters around $\log \teff = 3.5$ and $\log L/\lsun = 5$ at
solar metallicity. It is not clear what the dominant mass loss
mechanism for these stars is. Are they cool enough to develop
dust-driven winds or is mass loss simply driven by radiation pressure?

Table~\ref{tabl:massloss} shows a compilation of observational and
theoretical mass loss rates, for a combination of typical SAGB
parameters. Note, however, that these rates are not constant over
time, and that the variation itself during the AGB phase is important
for the final outcome. We preferentially use the observed mass loss
rates for massive AGB stars and red supergiants by
\citet[L05]{lcz+05}.  If dust-formation does not play an important
role, then the Reimers mass loss rate \citep{rei75}, may be
applicable. It is derived from observations of RGB stars with a small
range in temperatures and radii, however, \citet{sc05} have revised
the Reimers rate.  For the more massive RSG stars their new approach,
which also includes surface gravity, gives about three times larger
mass loss than the Reimers formula.  This places it, for given
temperature and luminosity (c.f.\ Table~\ref{tabl:massloss}), within a
factor of 2 of the observational mass loss determination by L05. The
mass loss formula by \citet{vassiliadis:93} (hereafter VW93), which is
often used for AGB star evolution calculations, is also close to the
observational value. The AGB mass loss formulated by \citet{blo95},
and based on the hydrodynamic wind models by \citet{bowen:88} has a
higher luminosity exponent, and gives very high mass loss rates for
SAGB stars.  From our first estimate in Sect.\ \ref{sec:sagb-simp_pop}
it is clear that with the Bloecker mass loss SAGB stars would never
explode as ECSN.

\begin{table}
\caption{Mass loss rates for SAGB stars with a typical value for
   the luminosity of $\log
   L/\Ls=5$ and an effective temperature of
   $T_{\mathrm{eff}}=3000$K.\label{tabl:massloss}}
\begin{tabular}{lrr}

 & Type & Rate\\ \hline
Reimers ($\eta=1$) & Red Giants & $\sim 5\cdot10^{-6}\msun$/yr\\
Reimers ($\eta=4$)  & Red Giants & $\sim 2\cdot10^{-5}\msun$/yr\\
Schr\"{o}der \& Cuntz & Super Giants & $\sim 1\cdot10^{-5}\msun$/yr\\
van Loon & AGB/RSG & $\sim 3\cdot10^{-5}\msun$/yr\\
Bl\"ocker & AGB & $\sim 6\cdot10^{-3}\msun$/yr\\
Vassiliadis \& Wood & AGB & $\sim 4\cdot10^{-5}\msun$/yr\\
\end{tabular}
\end{table}
\section{Results}
\label{sec:results}

We perform a series of synthetic calculations, with two assumptions on
third dredge-up and three assumptions on mass loss.  For dredge-up we
assume either the parameterization of \S~\ref{sect:lambda} or $\lambda
= 0$. For mass loss, we consider the cases Reimers, L05, and
VW93 (\S~\ref{sect:massloss}).

\subsection{Initial mass range for ECSN}

\begin{figure}
\includegraphics[angle=270,scale=0.31]{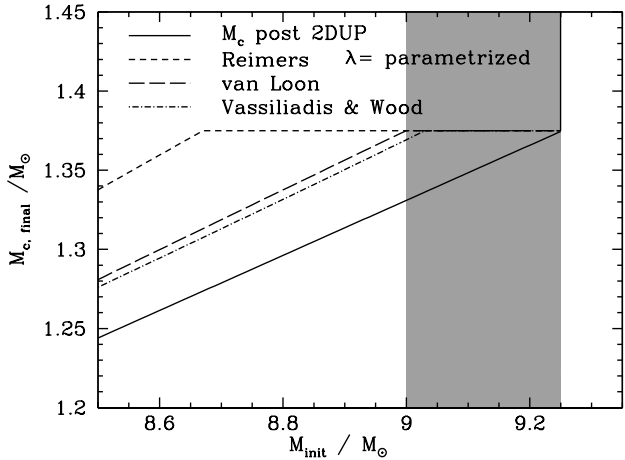}
\caption{Final core mass as a function of initial mass based on
   synthetic SAGB calculations. The solid line indicates the post
   second dredge-up core mass, the short dashed line indicates the
   final core mass using the Reimers mass loss rate ($\eta=4$), the
   dashed line using the L05 mass loss rate, and the dash-dotted line
   using the VW93 mass loss rate. The shaded region indicates the
   initial mass range for ECSNe for the L05 mass loss rate.}
\label{fig:Q_Lpara}
\end{figure}

\begin{figure}
\includegraphics[angle=270,scale=0.31]{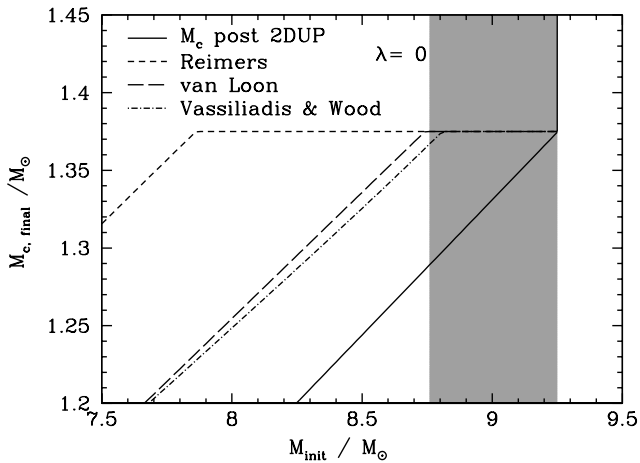}
\caption{Same as Figure~\ref{fig:Q_Lpara} but for calculations
   assuming no dredge-up. Note different x-axis scale.  }
\label{fig:Q_L0}
\end{figure}

The resulting initial mass ranges for ECSN are illustrated in
Figure~\ref{fig:Q_Lpara} for the case with parameterized $\lambda$,
and in Figure~\ref{fig:Q_L0} for $\lambda = 0$.  Stars that end their
evolution as white dwarf, i.e. below the Chandrasekhar mass, do not
explode as ECSNe.  With the parameterized prescription for the third
dredge-up, the width of the initial mass window for which ECSN occurs
is between 0.25\msun and 0.65\msun, depending on the assumed mass loss
rate. The mass loss prescriptions of L05 and of VW93 give an initial
mass window of $0.20-0.25\msun$.  With zero dredge-up, the core grows
at the maximum possible rate.  The width of the initial mass window
for ECSN is between $1.4\msun$ for the Reimers mass loss rate and
$0.45-0.5\msun$ for the VW93 and L05 mass loss rates.

\begin{table*}
\caption{Initial mass limits for ECSN and ratio of ECSN to SN as a
   function of the dredge-up efficiency and mass loss prescription.}
\label{tabl:SNrates}
\begin{center}
\begin{tabular}{l|ccc|ccc}
  & \multicolumn{3}{c|}{$\lambda =$ parameterized} &
\multicolumn{3}{c}{$\lambda =0$}\\[0.5em]
& $M_{\mathrm{low}}$/\msun & $M_{\mathrm{high}}$/\msun & \% EC
  & $M_{\mathrm{low}}$/\msun & $M_{\mathrm{high}}$/\msun & \% EC \\
\hline
Reimers ($\eta=4$)  &$ 8.67 $ & $9.25 $& $ 8.4   $ & $  7.86 $ & $9.25 $& $ 19.7 
$  \\
VW93      &$ 9.03 $ & $9.25 $& $ 3.2   $ & $  8.82 $ & $9.25 $& $  
6.2  $  \\
L05  &$ 9.00 $ & $9.25 $& $ 3.6   $ & $  8.76 $ & $9.25 $& $  
7.1   $ \\
\end{tabular}
\end{center}
\end{table*}

\subsection{ECSN fraction}

Based on the inferred mass ranges from the synthetic model, we
determine the ratio of the number of ECSNe to the total number of SNe.
Table~\ref{tabl:SNrates} gives an overview of the results for the
cases of parameterized dredge-up and without dredge-up ($\lambda=0$),
assuming the Salpeter IMF.  The value of $\lambda$ has a strong
influence on the predicted fractions.  With the parameterized
dredge-up and the VW93 or L05 mass loss rates the ECSN fraction of all
supernovae is about $3.5\%$.  With the Reimers mass loss rate $8\%$ of
all supernovae are ECSN.  The largest ECSN fraction of $20\%$ is
obtained without dredge-up and using the Reimers mass loss rate.

\subsection{Final masss and SN properties}

\begin{figure}
\includegraphics[angle=270,scale=0.31]{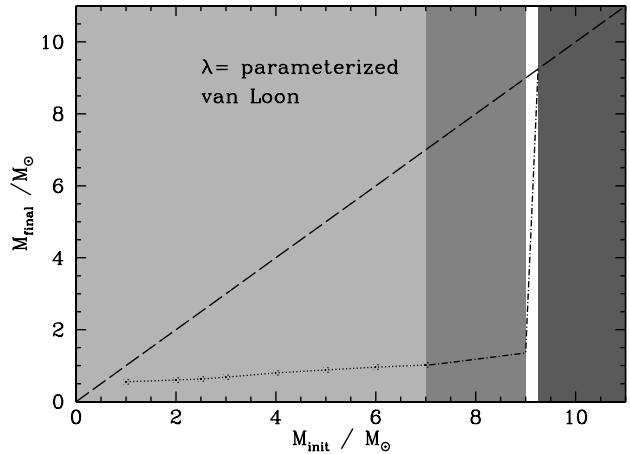}
\caption{Final mass of the remnant as a function of the initial mass.
   Remnant regimes are shaded as \textsl{light grey:} CO-white dwarf;
   \textsl{grey:} ONe white dwarf; \textsl{white:} ECSN; \textsl{dark
   grey:} CCSN.  The final mass is either the WD mass or the stellar
   mass at the time of SN explosion.  The dashed line indicates the
   line of initial mass equal final mass.}
\label{fig:ifrel}
\end{figure}

Figure~\ref{fig:ifrel} shows the initial-final mass relation for the
mass range from $1.0\msun$ to $14\msun$ using the parameterized
dredge-up prescription and the L05 mass loss rate.  For ECSNe, we find
a large spread in progenitor and envelope masses. The least massive
SAGB SN progenitors lose almost their entire envelope, growing the
core just barely enough to still make an electron capture supernovae
before the envelope is lost. The most massive SAGB SN progenitors, on
the other hand, undergo very little TP-AGB mass loss before they
explode, and contain a massive hydrogen-rich envelope at that time.

This diversity is a natural consequence of the competition between
core growth and mass loss during the SAGB stage, and thus independant
of the choice of mass loss rate and dredge-up parametrisation.  The
expected envelope mass range, from almost zero to about 8\msun
(\abb{fig:ifrel_zoom}), implies a diversity of supernova light curves
of ECSNe, which may range from light curves of so called Type~IIb and
Type~IIL supernovae to those of typical Type~IIP supernovae
\citep{fa77,you04}.

\begin{figure}
\includegraphics[angle=270,scale=0.31]{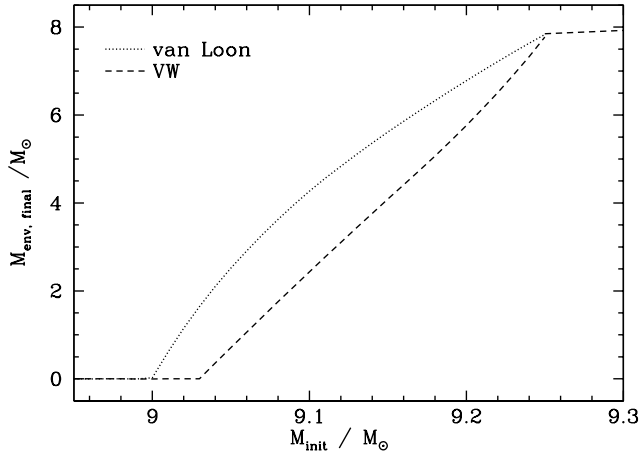}
\caption{Mass of the envelope just prior to the explosion as function
   of the initial mass for two mass loss rates. 
}
\label{fig:ifrel_zoom}
\end{figure}

\begin{figure}
\includegraphics[angle=270,scale=0.31]{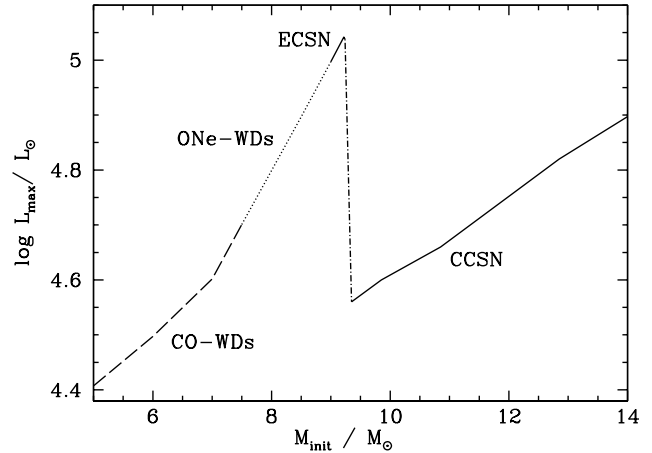}
\caption{Maximum luminosity as funtion of the initial mass at the end
of the evolution. Progenitors of CO white dwarfs reach luminosities up
to $\log L/\lsun \sim 4.6$, progenitors of ONe white dwarfs reach
luminosities up to $\log L/\lsun \sim 5$. Progenitors of ECSN are the
most luminous, with $\log L/\lsun \ge 5$, while progenitors of the
least massive CCSN have pre-explosion luminosities of $\log L/\lsun
\sim 4.6$}
\label{fig:maxlum}
\end{figure}

\begin{figure}
\includegraphics[angle=270,scale=0.31]{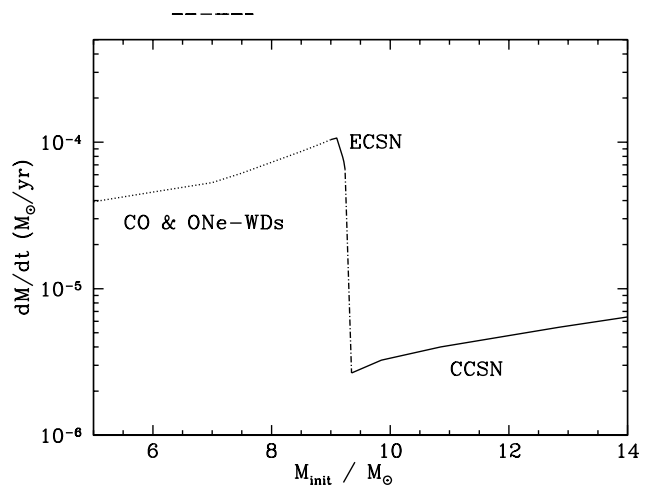}
\caption{Mass loss rates according to L05 as function of initial mass,
for the models shown in Fig~\ref{fig:maxlum}.}
\label{fig:finmdot}
\end{figure}

However, ECSNe may show three properties which might allow to
distinguish them from ordinary Type~II supernovae. First, they might
produce low-energy explosions \citep{kjh06} and possibly low neutron
star kicks \citep{plp+04}. A consequence of the low explosion energy
may be a small nickel mass produced by the explosion, and thus a low
luminosity of the tail of the light curve which is produced by the
decay of $^{56}$Ni and $^{56}$Co \citep{kjh06}.  The Type~IIP SN~1997D
may provide an example \citep{cu00}. Second, however, the enormous
mass loss rate of the supernova progenitor (Fig.~\ref{fig:finmdot})
star may produce clear signatures of a supernova-circumstellar medium
interaction in the supernova light. Such signatures are in particular
exceptionally a bright and long-lasting light curve \citep{skl01}, and
narrow hydrogen emission lines superimposed to a typical SN~II
spectrum \citep{ptb+02}.  Third, ECSN progenitors are extremely
bright, with luminosities of the order of $10^5\, L_{\odot}$
(Fig.~\ref{fig:maxlum}). Thus, progenitor identifications on
pre-explosion images \citep{ms05,hsc+06} might be able to identify
ECSNe.  They may be distinguished from very massive ($> 20\,
M_{\odot}$) progenitors of similar luminosity by their much cooler
effective temperatures ($<3000$K for ECSN progenitors versus $\sim
3400$K for CCSN progenitors).

\subsection{The reference model, examples}
\label{sect:pref_model}

As shown above, the results of our synthetic SAGB calculations
employing the VW93 and the L05 mass loss prescriptions are rather
similar. Since it is unclear whether the Reimers mass loss rate is
really applicable, and as the L05 mass loss rate relies on very recent
observations, we adopt the latter as the fiducial mass loss
prescription for our synthetic SAGB modeling.  Concerning the third
dredge-up efficiency, we adopt the mass dependant formulation shown in
Fig.~\ref{fig:3DUP} as our reference efficiency. These assumptions
define our reference model for the synthetic SAGB evolution.

\begin{figure}
\plotone{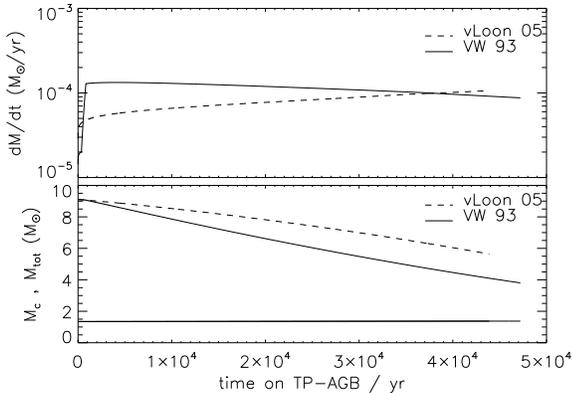}
\caption{Synthetic AGB mass evolution for an initial mass of
   $9.1\msun$. Upper panel: mass loss rate for two cases (L05 and
   VW93, \S~\ref{sect:massloss}); the superwind regime in the VW93
   mass loss prescription sets in at $1000\mem{yr}$ right at the
   beginning of the AGB phase. Lower panel: evolution of the core and
   the total stellar mass.}
\label{fig:TPev91}
\end{figure}

In the following, we discuss some explicit examples to illustrate the
TP-SAGB evolution, and to further motivate the choice of our reference
model and analyse its uncertainty.  The first example is a star with
an initial mass of $9.1\msun$, with a He-core mass at the end of the
second dredge-up of $1.348\msun$.

During the evolution on the TP-SAGB the luminosity first increases
from $\log L/\lsun = 4.9$ to $\log L/\lsun = 5.02$, and then drops
slightly due to decreasing envelope mass which decreases the
efficiency of hot-bottom burning.  As a result, the inter-pulse core
growth rate increases from $3 \times 10^{-7} \msun/\mathrm{yr}$ to
$1.5 \times 10^{-6} \msun/\mathrm{yr}$, but the effective core growth
is significant lower (by about factor 0.5) due to the effect of
dredge-up.  The mass loss rate increases from $3 \times 10^{-5}
\msun/\mathrm{yr}$ to $1 \times 10^{-4} \msun/\mathrm{yr}$
(Fig.~\ref{fig:TPev91}). In this model the SAGB ends after $\sim
4.4\times 10^{4} ~\mathrm{yr}$ when the core reaches its Chandrasekhar
mass.  The remaining envelope has a mass of $4.27\msun$.

For a $10\%$ larger dredge-up efficiency the SAGB time increases to
$\sim 4.8\times 10^{4} \mathrm{yr}$, and the remaining envelope
decreases to $3.82\msun$. For a $10\%$ smaller dredge-up efficiency
the SAGB decreases to $\sim 4.0\times 10^{4} \mathrm{yr}$, and the
final envelope mass increases to $4.61\msun$.  For the case of no
dredge-up the SAGB time is $\sim 2.4\times 10^{4} \mathrm{yr}$ and the
final envelope mass is $6.1\msun$.

Using the mass loss prescription of VW93 and the parameterized
dredge-up, the result is only different in the mass loss history,
resulting in a smaller final mass.  The main reason is that this mass
loss prescription accounts for the superwind phase for TP-AGB stars,
which gives a significantly different evolution of the envelope.  The
TP-AGB phase starts with low mass loss but makes a transition to the
superwind phase after $1000\jahre$ with mass loss rates around
$1\times 10^{-4}\msun/\jahre$, decreasing slowly due to the waning
luminosity (Fig.~\ref{fig:TPev91}, Upper Panel).  After $4.7\times
10^{4}\jahre$, the core reaches its Chandrasekhar mass with a final
envelope mass of $2.43\msun$.

\begin{figure}
\plotone{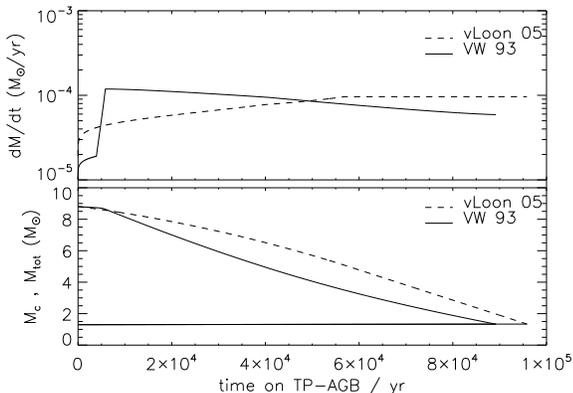}
\caption{Similar to Figure~\ref{fig:TPev91} but for a $8.8\msun$
star.  The core does not reach the Chandrasekhar
mass but loses its envelope and becomes a massive ONe
white dwarf.}
\label{fig:TPev88}
\end{figure}

An initially $8.8\msun$ star with the same mass loss and dredge-up
(Fig.~\ref{fig:TPev88}) becomes an ONe WD. The He-core mass at the
beginning of the SAGB is $1.296\msun$. During the SAGB evolution the
luminosity increases to $\log L/\lsun = 4.95$, then decreases due to
mass loss.  Assuming the L05 mass loss rate, the envelope is lost
after $\sim 1 \times 10^{5}\yr$. The mass loss rate increases steadily
during the TP-AGB phase, but as the star reaches a surface temperature
of $2500\mathrm{K}$ (Sect.~\ref{sec:sagb-synt}) we assume a constant
mass loss rate of $\dot{M} = 1 \times 10^{-4}\msun/\yr$ during the
last phase of the evolution. This is consistent with the observations
of L05 who find only one star with a higher mass loss rate.  The
remaining ONe core has a mass of $1.338\msun$. Repeating this
calculation with the mass loss rate of VW93 does not show large
differences (Fig.~\ref{fig:TPev88}).

\begin{table*}
\caption{ Final core and envelope masses for different initial
masses. The first three columns give the initial conditions of the
models. Models marked with an asterisk are calculated until the end of
the second dredge-up with EVOL (Table~\ref{tabl:overview}), the other
models are interpolated assuming a linear relation between the initial
mass and the core mass at the end of the second dredge-up. The fourth
and the fifth column give the final core and envelope mass for two
different mass loss prescriptions (\S~\ref{sect:massloss}) and for a
parameterized dredge-up efficiency (\S~\ref{sect:lambda}). The sixth
and seventh column give the final core and envelope mass for a
dredge-up efficiency of zero. In case a column contains ''WD'' or
''NS'', the remnant is respectively a white dwarf or a neutron
star. All masses are in units of a solar mass.}
\begin{tabular}{lll|lr|lr|lr|lr}\\
\multicolumn{3}{c}{ initial
   conditions}&\multicolumn{4}{|c}{ final conditions
   param. $\lambda$}& \multicolumn{4}{|c}{ final
   conditions  $\lambda = 0$}\\
\multicolumn{3}{c}{}&\multicolumn{2}{|c|}{\itshape
   L05}&\multicolumn{2}{c}{\itshape VW 93}&\multicolumn{2}{|c|} 
{\itshape L05}&\multicolumn{2}{c}{\itshape VW 93}\\ \hline
$M_{\mathrm{tot}}$ & $M_{\mathrm{c}}$ & $M_{\mathrm{env}}$ &
   $M_{\mathrm{c}}$ & $M_{\mathrm{env}}$ & $M_{\mathrm{c}}$ &
   $M_{\mathrm{env}}$&
   $M_{\mathrm{c}}$ & $M_{\mathrm{env}}$ & $M_{\mathrm{c}}$ &
   $M_{\mathrm{env}}$\\ \hline
  6.5000* &  0.9900 &  5.5100 &  1.0039 &   WD    &  1.0060 &  WD    &
   1.0816 &   WD   &  1.0887 &   WD        \\
  7.5000* &  1.0700 &  6.4300 &  1.0886 &   WD    &  1.0903 &  WD    &
   1.1734 & WD  &  1.1739 & WD    \\
  8.5000* &  1.2440 &  7.2560 &  1.2809 &   WD    &  1.2761 &  WD     
&  1.3358 & WD &  1.3255 & WD   \\
  8.7500 &  1.2875 &  7.4625 &  1.3282 &   WD    &  1.3223 &  WD      
&  NS &  0.2075 &  1.3654 & WD   \\
  8.8000 &  1.2962 &  7.5038 &  1.3376 &   WD    &  1.3316 &  WD      
&  NS &  1.3774 &  1.3736 & WD   \\
  8.9000 &  1.3136 &  7.5864 &  1.3563 &   WD    &  1.3504 &  WD      
&  NS &  3.2631 & NS  &  1.4514    \\
  9.0000 &  1.3310 &  7.6690 &    NS   &  0.0221 &  1.3693 &  WD      
&  NS &  4.7723 & NS  &  3.0680    \\
  9.1000 &  1.3484 &  7.7516 &    NS   &  4.2672 &    NS   &  2.4311  
&  NS &  6.0628 & NS  &  4.7208 \\
  9.2000 &  1.3658 &  7.8342 &    NS   &  6.7800 &    NS   &  5.7643  
&  NS &  7.2711 & NS  &  6.6314\\
  9.2500 &  1.3745 &  7.8755 &    NS   &  7.8321 &    NS   &  7.7731  
&  NS &  7.8481 & NS  &  7.8138 \\
\hline
\end{tabular}
\label{tabl:final_overview}
\end{table*}

\section{Concluding remarks}
\label{sec:disc}
We show that both the lower initial mass for SAGB stars and the
maximum initial mass for ECSNe sensitively depend on the assumptions
for mixing during core H- and He-burning.  EVOL models which include
core overshooting, and which are consistent with the observed width of
the main sequence, predict a smaller maximum initial mass for SAGB
stars.  Rotation would act similar to overshooting during the core
burning phases.  The STERN models include neither rotation nor
overshooting, and the maximum initial mass for SAGB and ECSNe is
larger by up to $2.5\msun$.  Equally important is the treatment of
semiconvection during He-core burning.

On the other hand, the lower initial mass limit for ECSNe is determined
by the stellar properties on the SAGB.  Most important are the
third dredge-up efficiency, the mass loss rate, and the hot-bottom
burning efficiency and its dependence on the adopted convection theory
for the envelope.  In general, larger mass loss, larger dredge-up
efficiency and large hot-bottom efficiency all decrease the initial
mass range for ECSNe, or even suppress the ECSN channel.  In order to
increase the accuracy of the transition initial mass between ECSNe and
CCSNe and of the lower mass limit for ECSNe, these classical issues of
stellar evolution need to be improved specifically for the initial
mass range of $6$ to $12\msun$.

We have discussed here the SAGB stars with C-ignition and formation of
ONe cores as the most likely progenitors of an ECSN class of
supernova. It is in principle possible that initially less massive
stars that develop CO cores could increase their core size on the
TP-AGB up to the Chandrasekhar mass resulting in a supernova
explosion.  Despite the uncertainties still involved we can rule out
this possibility for solar metallicity. There are two main reasons
that prevent these SN1.5a from occurring: First, the mass loss would
have to be much lower than observed. Second, models predict that the
third dredge-up is larger for massive AGB stars with initial mass
between $4$ and $7\msun$ than for SAGB stars
\citep[\abb{fig:3DUP};][]{klp02}. This makes it even more unlikely for
massive AGB stars to significantly grow their cores.

We did not take into account mass loss until the beginning of the
thermally pulsing phase. If mass loss were applied during the main
sequence and up to the TP-AGB, less than half a solar mass would have
been lost (\citealt{sie06}, their Table 5).  This may shift the quoted
initial masses to a slightly higher value.

We note that there is large disagreement between the different studies
on the dredge-up efficiency in SAGB stars \citep{rgi96b, dl06, sp06}.
Whereas \citet{rgi96b} and \citet{sp06} find negligible amounts of
dredge-up, \citet{dl06} find very efficient dredge-up with $\lambda
\sim 0.7$. This situation is similar to the divergent modeling results
obtained in the past on the third dredge-up in low-mass AGB stars. For
the low-mass regime there is now some consensus that most of the
differences where related to different physical and numerical
treatments of the convective boundaries. Possibly the same applies to
the divergent results for SAGB stars. In our synthetic model we
adopted a parameterized prescription (\abb{fig:3DUP}) that is based on
state-of-the-art full stellar evolution calculations. To test the
effect of dredge-up we also considered a case with no dredge-up.
Clearly, the third dredge-up in SAGB stars needs to be studied in more
detail. It is closely related to the mixing conditions at the bottom
of the convective boundary. This is a hydrodynamic situation which
requires multi-dimensional simulation which is complicated by the fact
that for these extremely massive cores the dredge-up seems to be hot
\citep{her04b}, i.e., any small amount of H that could be mixed across
the convective boundary will instantly burn violently with all the
associated feedback on the evolution of the convective instability in
that region. On the other hand, the amount of material that is
dredged up from the He intershell will be very small, even with large
values of $\lambda$, due to the thin intershell, and will therefore
easily dilute in the envelope.

It is presently not known whether ECSNe from SAGB stars contribute to
the \textsl{r}-process pattern in the universe.  Explosions from stars
in this mass range have been investigated as promising site for the
astrophysical \textsl{r}-process (\citealt{wch98,stm+01, wti+03b}),
based on the work of \citet{hwn84} who exploded a ONe core model from
\citet{nom84e,nom87c}.  Other groups were not able to confirm the
\textsl{r}-process contribution due to the low entropy (e.g.,
\citealt{bl85, bck87, mw88, bw85}). \citet{kjh06} ruled out this
possibility based on updated physics and two different nuclear
equations of state.\\

In any case, our study outlines that ECSNe from SAGB stars are likely
to occur, if only at a level of a few percent of the Type~II
supernova rate in the local universe. However, at low metallicity,
the key physical ingredients to the evolution of thermally pulsing
SAGB stars may change. In particular, the stellar wind mass loss rate
may be lower, which might open the ECSN channel appreciably, and may
even allow Type~1.5 supernovae. This issue will be discussed in
a forthcoming paper.


\acknowledgments 
This work has been in part supported by the
Netherlands Organization for Scientific Research (NWO) (AJP). AH and FH
are supported under the auspices of the National Nuclear Security
Administration of the U.S. Department of Energy at Los Alamos National
Laboratory under Contract No. DE-AC52-06NA25396.  This research was
supported by the DOE Program for Scientific Discovery through Advanced
Computing (SciDAC; DOE DE-FC02-01ER41176 and DOE DE-FC-02-06ER41438).

\bibliographystyle{apj}

\end{document}